# Broadband, compact, and training-free optical processors for parallel image classification

Sander J. W. Vonk, Boris de Jong, Yannik M. Glauser, David B. Seda, Matthieu F. Bidaut, Benjamin Savinson, Hannah Niese, and David J. Norris*

Optical Materials Engineering Laboratory, Department of Mechanical and Process Engineering,
ETH Zurich, 8092 Zurich, Switzerland.

**As artificial intelligence becomes increasingly prevalent, the demand for faster and more energy-efficient computing approaches grows. While optical computing offers intrinsic advantages in bandwidth and power consumption, existing implementations remain bulky, wavelength-specific, and dependent on complex training procedures, limiting scalability and parallel operation. In this work, we demonstrate a compact, training-free optical processor based on wavy diffractive features, known as Fourier surfaces, for parallel image classification. Our device achieves classification accuracies of up to 84% for digit datasets and 66% for fashion datasets within a 40×40 µm$^2$ footprint. The diffractive layer inherently separates incident wavelengths into distinct output directions, enabling broadband operation and allowing multiple colors to function as independent computation channels. As a result, this passive system supports up to 20 simultaneous computations within a single optical pass. These results highlight the potential of nanoscale diffractive systems to achieve high compute densities, paving the way for scalable, low-power optical processors for machine learning and image-recognition applications.**

## Introduction

As artificial intelligence (AI) systems continue to grow in complexity and size, their power demands are emerging as a critical bottleneck.[1] Energy consumption of state-of-the-art AI models is already straining data infrastructure and depleting resources. Alternative computing architectures are therefore needed that can deliver high performance at drastically reduced energy costs.

Optical computing offers a promising solution due to the inherently high bandwidth and low-loss nature of light.[2,3] Optical systems can perform computations with far greater energy efficiency than electronic processors,[2] with recent demonstrations even approaching sub-photon energy usage per operation.[4] Research in optical computing has therefore accelerated, spanning diffractive and scattering-based approaches aimed at overcoming the limitations of electronics.[5-16] However, many of these implementations remain bulky and rely on large optical components such as spatial-light modulators or extended cascaded layers, restricting their scalability. Furthermore, the design and training of optical networks pose significant challenges,[17,18] as optimizing the device parameters for a given computational task often requires large training datasets and iterative optimization. Achieving broadband operation adds an additional layer of difficulty, since the performance of diffractive and scattering-based systems is often highly wavelength-dependent and typically requires deep-learning-assisted design to extend functionality across multiple wavelengths.[19,20] These challenges have motivated the search for compact, easily trainable optical architectures capable of achieving high computational density and broadband operation.

Here, we design, fabricate, and benchmark an optical processor based on wavy diffractive structures, known as Fourier surfaces, for parallel image classification.[21] This approach has only recently become feasible due to the continuous height control enabled by thermal scanning probe lithography (TSPL), which allows the fabrication of smoothly varying, wavy diffractive surfaces rather than discretized or binary patterns. As such, the device operates without iterative training: the surface height profile is determined from training datasets in a single analytical step, without gradient-based optimization or backpropagation. The resulting diffractive features redirect optical image inputs into class-specific output directions. By operating directly on optical image inputs, the processor avoids intermediate electro–optical conversion, reducing latency and energy consumption. Each output channel corresponds to a characteristic feature pattern derived from the image dataset, enabling intensity-based classification. Because the diffracted outputs for a single color are directed to distinct polar angles, the system operates inherently broadband, allowing different wavelengths to function as independent computational channels. Theory shows that up to 20 wavelengths can be resolved simultaneously, and experiments confirm robust all-optical inference for multiplexing factors up to six. Overall, our single-layer design achieves classification accuracies of up to 84% for handwritten digits and 66% for fashion items within a compact 40×40 µm$^2$ footprint, while enabling multi-wavelength parallel processing—a combination that highlights its potential for scalable, low-power, and high-throughput photonic computing.

## Results and discussion

### Training-free optical classification in Fourier space

Figure 1a schematically shows the working principle of the single-layer optical processor in silver. We project an optical input—in this example an amplitude image of a zero from the digit MNIST dataset[22]—onto the wavy diffractive surface. The diffractive surface introduces local optical path length differences, imprinting a spatially varying phase onto the incident amplitude input. This amplitude and phase profile diffracts into predefined directions that are collected by our microscope objective. We image the back focal plane (Fourier space) of our microscope objective to directly map the diffraction angles of the optical processor. A high optical intensity in the designated output port leads to a successful classification of the input image.

The design of the optical processor is entirely training-free, and its wavy structure is built up from diffractive grating structures that serve as class-specific spatial templates. As a simple example, Figure 1 shows the design for a binary classifier for input images of zeros and ones. For both classes in this dataset, we compute the average spatial distribution (Figure 1b,c; all zeros and ones in the training set), which acts as a spatial weighting of sinusoidal gratings with a pitch $p = $ 1 µm and a unique diffraction direction. In this case, a grating into the $x$-direction for the one and the $y$-direction for the zero (Figure 1b,c; insets show the two-dimensional Fourier spectrum). Superposition of these weighted gratings forms the design for a single diffractive layer that maps distinct amplitude images to specific angularly separated diffraction orders, or output ports.

For the fabrication, the wavy design (Supporting Information Figure S1) is first patterned into a thermo-responsive polymer using TSPL.[21,23] Here, the maximum depth of the structure is fixed to 150 nm. After patterning, silver is evaporated onto the polymer[24] and finally, the optical processor is obtained through template stripping.[25] Figure 1d shows a scanning electron micrograph (SEM) of the optical processor in silver.

For binary inference, we illuminate the device with input test images at a wavelength $\lambda = $ 650 nm using a phase-only spatial-light modulator (SLM). By placing the SLM in the illumination path between crossed polarizers, we achieve images at the device via amplitude modulation (see Methods). Figure 1e,f shows examples of recorded real-space images of a projected zero and one reflected from flat silver. For this, the 28×28 pixel images are digitally upscaled to cover 960×960 SLM pixels (7.7×7.7 mm$^2$), such that the image of the SLM plane matches the size of the reflective layer (40×40 µm$^2$) after propagating through our optical setup (see Methods). The spatial overlap of the input image with the corresponding grating template maximizes the diffracted intensity in the correct output direction (Figure 1e,f; insets for corresponding measurements of Fourier space), enabling optical classification. The bright spot in the center of Fourier space is the specular reflection of the input image. For each measurement $n$, we integrate the intensity over both the $+1$ and $-1$ diffraction orders for each output channel to obtain the raw optical output $\mathbf{x}^{(n)}$. Then, we renormalize by the average intensity per output channel over all measurements $\bar{\mathbf{x}}$ to obtain the normalized optical output $\tilde{\mathbf{x}}^{(n)} = \mathbf{x}^{(n)} \oslash \bar{\mathbf{x}}$, with $\oslash$ representing element-wise division. This renormalization compensates for differences in the average spatial extent of images across input classes; for example, zeros occupy a larger area on average than ones. Theoretically, light can also be diffracted by higher-order diffraction processes, but these have low efficiencies for our shallow diffractive surfaces (see Supporting Information Figure S2). Figure 1g shows the output intensities in the two output channels for all zeros (blue dots, 980 inputs) and ones (red dots, 1,135 inputs) from the digit MNIST test set. A successful classification has the highest intensity in the designated output port, so $y > x$ (Figure 1g, black line gives $y = x$) for a zero and $y < x$ for a one. We achieve classification accuracies of 99.9% and 99.8% for zeros and ones, respectively (Figure 1g, inset).

**Full digit classification**

Building upon the binary optical classifier, we extend the training-free concept to full digit classification by simply encoding ten distinct sinusoidal gratings within a single reflective silver layer. Each grating corresponds to a digit class (0–9) and is oriented at a unique azimuthal angle $\varphi$. The angles are spaced by $\Delta\varphi = \pi/10$, such that each class diffracts light into a distinct output direction and the optical outputs are evenly distributed in Fourier space. The superposition of these gratings forms a compact diffractive surface that again directs optical inputs to class-specific output ports based on their spatial overlap with the corresponding spatial template. Figure 2a shows an SEM of the fabricated optical processor in silver. Visually, overlapping features from different digit classes superpose their gratings, producing regions of enhanced structural symmetry reminiscent of quasicrystalline order.[21] The optical output in Fourier space (Figure 2b) for an input image of a zero, clearly shows that the diffracted intensity is maximal into the two output directions assigned to the zero class, consistent with a successful classification. For visualization, the intensity is clipped to $0.1\ I/I_{\max}$, as the fixed 150-nm depth reduces the diffraction efficiency of each individual sinusoid in the full ten-class device.

To quantitatively validate the performance of the full digit classifier, we perform a benchmark using all $N = 10{,}000$ test images from the digit MNIST dataset. For input $n$, the optical response vector $\tilde{\mathbf{x}}^{(n)}$ is obtained using the same readout/normalization procedure as for the binary classification (Figure 1), but now with 10 output classes with labels $t^{(n)} \in [0,9]$. The predicted class label is then determined as $\hat{t}^{(n)} = \arg\max \tilde{\mathbf{x}}^{(n)}$, where $\arg\max(\cdot)$ returns the element (or output port) of $\tilde{\mathbf{x}}^{(n)}$ with the highest intensity. From these measurements, we construct the confusion matrix

$$P_{i,j} = \frac{\sum_{n=1}^{N} \delta_{t^{(n)},i} \delta_{\hat{t}^{(n)},j}}{\sum_{n=1}^{N} \delta_{t^{(n)},i}}, \tag{1}$$

which gives the fraction of the inputs of true class $i$ being interpreted as class $j$ (Figure 2c). Here, $\delta_{i,j}$ denotes the Kronecker delta, equal to 1 when $i = j$ and 0 otherwise. The histogram as an inset shows the single-digit classification accuracy $P_{i,i}$. The overall classification accuracy $\eta$ is obtained from the diagonal elements of the confusion matrix as

$$\eta = \frac{1}{N}\sum_{n=1}^{N} \delta_{t^{(n)},\hat{t}^{(n)}} , \qquad (2)$$

yielding an accuracy of $\eta = 76\%$ for full all-optical classification across ten digit classes.

To boost the classification accuracy with minimal additional electronic computational effort, we introduce a simple linear correction step applied to the optical outputs $\tilde{\mathbf{x}}^{(n)}$ (Figure 2d). More specifically, we train a 10×10 matrix $\mathbf{M}$ that performs a linear transformation on the measured intensity vectors,

$$\tilde{\mathbf{y}}^{(n)} = \mathbf{M}\,\tilde{\mathbf{x}}^{(n)}, \qquad (3)$$

such that the transformed outputs $\tilde{\mathbf{y}}^{(n)}$ maximize the overall classification accuracy. This post-processing step can be interpreted as a calibration of the optical response, compensating for likely channel crosstalk between sets of output ports. Figure 2d presents the resulting confusion matrix and single-digit accuracy histogram obtained using a matrix $\mathbf{M}$ trained on the full set of measured optical responses. The classification accuracy increases from 76% to 84%, demonstrating that this simple electronic transformation can effectively enhance the performance of the optical classifier without adding significant computational overhead. To verify that this procedure learns a meaningful transformation rather than overfitting, we additionally trained $\mathbf{M}$ on half of the optical outputs and evaluated the classification accuracy on the remaining held-out half, yielding comparable performance as for training on the full output [82% (training) versus 83% (test)].

Intermediate devices between binary and full ten-digit classifiers are presented in Supporting Information Figure S1 (SEMs and height-profile designs) and Supporting Information Figure S3 (confusion matrices from theory and experiment). These results illustrate the progressive scaling of the classification accuracy as the number of encoded classes increases. The all-optical classification accuracies range from 100% for the binary device to 76% for the ten-class implementation, while application of the linear correction matrix **M** improves the accuracies to between 100% and 84%. All experimental results show good agreement (within a few percentage points) with predictions from simple scalar-diffraction theory, as explained in the Supporting Information Section S1.

**More challenging input data: fashion items**

To further test the versatility of our optical-computing concept, we extend it to a more challenging classification task involving fashion items from the fashion MNIST dataset.[26] This dataset comprises ten clothing categories such as shoes, T-shirts, and dresses (Figure 3a). In contrast to handwritten digits, these objects exhibit far greater inter-class similarity and overlapping spatial features, making them a more demanding benchmark for optical inference. We design and fabricate a new diffractive structure using the same principles as before, encoding ten sinusoidal gratings within a single reflective silver layer. Again, each grating corresponds to one fashion category and is oriented at a distinct azimuthal angle $\varphi$. The resulting device has a maximum depth of 150 nm and a lateral footprint of 40×40 µm$^2$ (Figure 3b for SEM).

Evaluation of the device using the complete fashion MNIST test set with 10,000 fashion items yields a classification accuracy of $\eta = 66\%$ when combining the optical outputs with the same linear optimization procedure applied to the digit classifier (Figure 3c). Although this accuracy is lower than for the digit dataset, it still exceeds random guessing by more than a factor of six, demonstrating that substantial feature extraction and class separation occurs in the optical domain. The confusion matrix shows expected difficulties between visually similar categories such as shirts, T-shirts, pullovers, and coats, reflecting the intrinsic ambiguity of the dataset rather than device limitations. Intermediate devices with fewer encoded classes, along with their corresponding SEM, height-profile designs, and confusion matrices, are provided in the Supporting Information Figures S4 and S5, respectively.

**Broadband operation enabling parallel computation power**

To highlight the intrinsic parallelism of our diffractive classifier, we demonstrate that the single-layer digit classifier from Figure 2 can operate in parallel at multiple wavelengths. The key idea is to encode different digit classes not only into distinct azimuthal diffraction angles, as before, but also into different illumination colors spanning the visible spectrum (Figure 4a). Because the classifier diffracts each wavelength into a ring in the back focal plane (Fourier space) with a radius proportional to the normalized in-plane wavevector $k_\parallel/k_0$, where $k_0 = 2\pi/\lambda$, inputs at different colors naturally separate into concentric circles (Figure 4b). As a result, each wavelength functions as an independent computational channel: a red input encodes one digit and appears at a large diffraction radius, whereas blue inputs map to smaller radii. This wavelength–momentum mapping allows the device to execute many independent classifications simultaneously without crosstalk, limited only by the wavelength spacing $\Delta\lambda$ and the numerical aperture (NA) of the collection optics.

To theoretically estimate the limits of wavelength-multiplexed operation, we first analyze the angular width of the diffracted orders in Fourier space as a function of the classifier size. For a single sinusoidal grating that spans the entire device area, scalar diffraction theory predicts a $L^{-1}$ scaling of the spot width (Figure 4c; dashed line), with $L$ the sidelength of the device. However, each handwritten digit occupies only a fraction of the total area $L^2$, producing broadened diffraction orders. To capture this effect, we simulate the average Fourier-space response of 10,000 digit-MNIST test images for different $L$ and extract the corresponding full-width-at-half maximum (FWHM; Figure 4c; data points). These data points still follow a $L^{-1}$ scaling, but with a larger prefactor (Figure 4c; solid line). Using this dataset-averaged FWHM, we calculate the maximum achievable multiplexing factor (MF) for a given NA, defined as MF = NA/FWHM (Figure 4d). For our device dimensions, this analysis suggests an upper limit of roughly 20 independent wavelength channels that could, in an idealized limit, be resolved simultaneously within the available angular range. Experimentally, however, the usable wavelength range is constrained. The reflectivity of silver decreases towards the near-ultraviolet, while at longer wavelengths the accessible Fourier space is limited by the numerical aperture of the objective, such that diffraction outputs for wavelengths approaching ~800 nm lie at the edge of the collection cone. As a result, the practical multiplexing factor in our system is determined by both material considerations and collection-angle limitations.

We experimentally validate wavelength-multiplexed operation by evaluating the classifier at increasing levels of spectral parallelism, using parallel classification at 3, 6, and 12 different illumination colors (Figure 4e–g). Here, parallel operation is emulated by reconstructing multi-wavelength inference in post-processing from independently measured single-wavelength responses (laser linewidth of ~1.5 nm), with inputs randomly selected across digit classes to mimic simultaneous operation. For threefold multiplexing (450, 550, and 650 nm), the diffraction orders of all 10,000 test inputs remain cleanly separated in Fourier space (Figure 4e, all measurements superposed). Increasing to a sixfold set of wavelengths (425–675 nm in 50-nm steps) still produces clearly distinguishable concentric diffraction rings, consistent with the theoretical multiplexing limit derived above. However, when extending to twelve wavelengths spanning 425–700 nm in 25-nm steps, the diffraction orders overlap. The resulting classification accuracies (Figure 4h) reflect the progressive degradation with increasing spectral crowding: for threefold multiplexing the average classification accuracy remains high at 68% all-optical and 76.8% after linear correction; for sixfold multiplexing we obtain 65.8% and 73.3%; while for twelvefold multiplexing the performance drops to 51.4% and 53.9%, respectively. Notably, when all twelve measurement sets at different wavelengths are evaluated individually, their average classification accuracy (67.7% and 78.5%) remains comparable to the single-wavelength operation in Figure 2, indicating that performance degradation under multiplexed operation arises from diffraction-order overlap rather than a loss of intrinsic device performance for different wavelengths.

The combination of a compact footprint and intrinsic wavelength-multiplexed parallelism enables our diffractive classifiers to reach compute densities, i.e., multiplications per second per square meter, beyond those accessible in digital processors.[27,28] Each device occupies only 40×40 µm$^2$, corresponding to a density of $10^{10}$ m$^{-2}$ computing units that can all process multiple inputs simultaneously through spectral separation. Experimentally, we demonstrate a limit of using 6–12 independent wavelengths, while our theoretical analysis indicates that up to 20 channels are feasible for this architecture. Each optical classification corresponds to a matrix–vector multiplication of a 28$^2$-dimensional input onto 10 output channels, amounting to ~10$^4$ physically constrained analog multiplications per inference. When paired with emerging ultrafast optical modulators, such as the LiNbO$_3$ spatial-light modulator achieving input rates approaching $10^9$ s$^{-1}$, each optical classifier could perform $10^{14}$ multiplications per second.[29] Scaled to a square meter of such devices, this corresponds to a compute density on the order of $10^{24}$ mathematical operations per second per square meter. Even with conventional commercial SLMs operating at ~$10^2$ Hz, the achievable compute density ($10^{17}$ s$^{-1}$ m$^{-2}$) remains competitive with electronic processors. These estimates highlight the potential of compact diffractive photonic systems to deliver improvements in scalable, low-power, high-throughput computation.

## Conclusions

In summary, we have introduced a compact, training-free optical computing platform that performs image classification using a single diffractive layer with a depth variation of only 150 nm. By encoding class-specific spatial-template gratings into a 40×40 µm$^2$ silver surface, our approach enables optical processing and classification with accuracies up to 84% for handwritten digits and 66% for fashion items. We demonstrated that the same device could operate as a parallel processor by assigning distinct computational channels to different illumination wavelengths. Theory predicts that up to 20 independent channels can be resolved, and experiments confirm robust performance for multiplexing factors up to six. When combined with emerging ultrafast modulators, these devices can reach compute densities of $10^{24}$ s$^{-1}$ m$^{-2}$. Our results highlight the potential of diffractive photonic architectures for scalable, low-power, high-throughput processing, capable of revolutionizing future computational hardware.

## Methods

**Fabrication of optical processors.** The single-layer optical processors were prepared using a 1-mm-thick, 2-inch-diameter Si (100) wafer (Silicon Materials). For this, the Si wafers were cleaned using oxygen plasma (GIGAbatch, PVA TePla) at 600 W for 2 min. A 12 wt.% solution of poly(phthalaldehyde) (PPA, Allresist) in anisole (AR 600-02, Allresist) was prepared and 300 µL was spin-coated onto the wafer using a two-step process. First, 5 s rotation at 500 rpm using a ramp rate of 500 rpm s$^{-1}$. Then, 40 s rotation at 2000 rpm with a ramp rate of 2000 rpm s$^{-1}$. The resulting film was patterned using a NanoFrazer Explore TSPL system (Heidelberg Instruments). For this, custom designs were loaded into the tool, and a depth of 150 nm was chosen. Patterning was performed by scanning a heated cantilever tip over the surface, locally sublimating the PPA. The depth at each pixel was controlled by modulating the electrostatic force between the cantilever and the substrate, allowing precise adjustment of the applied downward force.

An optically thick silver film (>500 nm) was deposited onto the patterned PPA later via thermal evaporation (Nano 36, Kurt J. Lesker). High-purity silver pellets (1/4-inch diameter × 1/4-inch length, 99.999%, Kurt J. Lesker) were used with a deposition rate of 25 Å s$^{-1}$ under a vacuum of approximately 3×10$^{-7}$ mbar. Following evaporation, a 1-mm-thick glass microscope slide (Paul Marienfield) was fixed to the silver surface using UV-curable epoxy (OG142-95, Epoxy Technology). To minimize residual PPA transfer to the final structure, the glass slide was allowed to settle onto the epoxy for ~5 min prior to UV curing, which was performed afterwards for 2 h. Finally, the glass–epoxy–silver stack was mechanically separated from the substrate using a razor blade, revealing the silver surface structure.

**Optical measurements.** The classification measurements were conducted by placing the optical processors on an inverted Nikon microscope (Nikon Eclipse Ti-U) equipped with a 50× air objective (Nikon TU Plan Fluor, NA = 0.8). A supercontinuum laser (SuperK Fianium, NKT Photonics) combined with a tunable band-pass filter (LLTF Contrast, NKT Photonics) provided illumination across a wavelength range of 400–1000 nm. The laser light was guided through an optical fiber (A502-010-110, NKT Photonics), then collimated using a 10× objective (Nikon TU Plan Fluor, NA = 0.3). It then passed through a 750-nm short-pass filter (FESH0750, Thorlabs) to eliminate parasitic light. Subsequently, the light was sent through a broadband 90:10 beam splitter (BSN10R, Thorlabs) to reflect 10% of the light to a power meter. The light from the supercontinuum laser (randomly polarized) was polarized with a linear polarizer (WP25M-VIS, Thorlabs) oriented at 45° relative to the horizontal axis of the SLM to only transmit light with 45° polarization angle. Then, the light goes through a visible 50:50 beam-splitter cube (BSW10R, Thorlabs) before reaching the reflective spatial light modulator (SLM, HOLOEYE PLUTO NIR-11). Amplitude modulation was achieved by placing a second linear polarizer at −45° behind the beam splitter. The amplitude profile is focused on the back focal plane of the 50× air objective by using a defocusing lens with focal distance of 75 cm (ACT508-750-A-ML, Thorlabs) resulting in a demagnification factor of 188. After reflecting off a broadband 50:50 beam splitter (AHF analysentechnik), the amplitude profile was directed to the objective, resulting in a Gaussian illumination profile across the 40×40 µm$^2$ substrate. The diffracted light from the optical processor, which was positioned at a real plane of the imaging system, was collected in reflection. The reflected light passed through a sequence of lenses (tube lens, Nikon; AC254-200-A-ML, Thorlabs; AC254-200-A-ML, Thorlabs; AC508-200-A-ML, Thorlabs), mirrors, and an iris, before being detected by a digital camera (Zyla PLUS sCMOS, Andor) located in a Fourier plane. The iris, placed in a real plane, was adjusted to collect light exclusively from the classifier.

To ensure optimal performance and reproducibility of the diffraction-based classification, we first verified that the amplitude profile enters the objective at normal incidence by aligning the specular reflection to the center of Fourier space. We optimized the dark–light contrast for the amplitude modulation at various wavelengths by changing the voltage range in the SLM.

We used two standard datasets for machine learning available through the PyTorch module: the digit MNIST (handwritten digits) and fashion MNIST (clothing items) datasets. Both datasets consist of grayscale images of 28×28 pixels. The images were binarized by applying a thresholding step: pixels were set to 1 if their intensity exceeded 0.5 for digit MNIST, and 0.1 for fashion MNIST.

# Author Information

**Corresponding Author.**  *Email: dnorris@ethz.ch

**Acknowledgements** This work was supported by the Swiss National Science Foundation (SNSF) through Award No. 2000-1-240090. S.J.W.V. acknowledges support from the Swiss National Science Foundation (SNSF) under Award No. 200021-232257 and from the ETH Career Seed Award. M.F.B. acknowledges support provided by a fellowship from the German Academic Exchange Service (DAAD). B.S. thanks ETH Zurich for support under Research Grant 25-1 ETH-030.

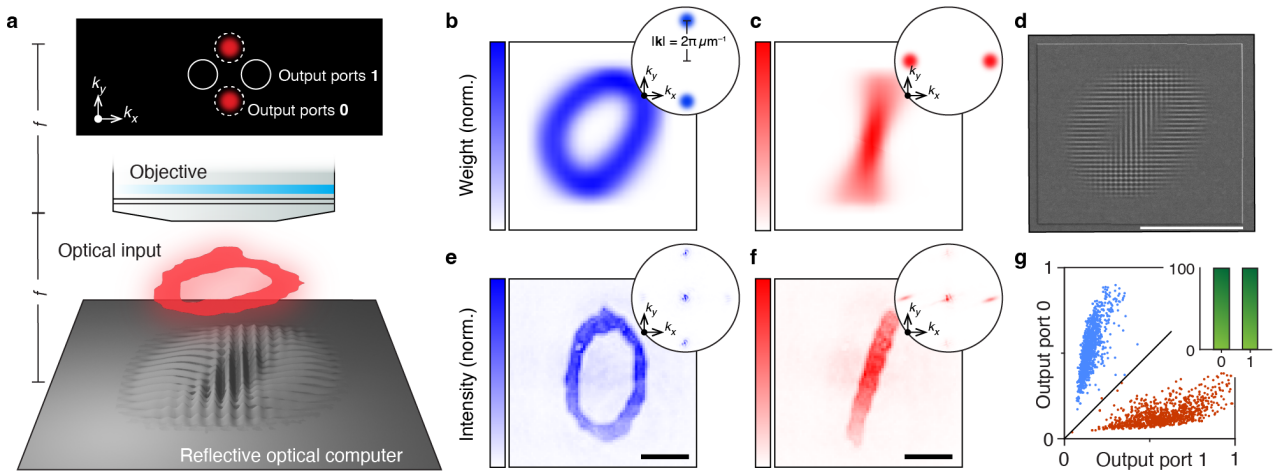

**Figure 1.** Training-free optical classification. (a) Schematic of the working principle of the compact reflective optical processor. Optical input images (encoded in amplitude) from the digit MNIST test set are projected onto the wavy diffractive surface. The resulting diffraction propagates to the back focal plane (Fourier space) of the microscope objective. Comparing the integrated intensities in all output ports enables classification of the optical input. (b,c) Average spatial distributions of the digit classes zero and one, used as spatial weights for constructing class-specific sinusoidal gratings with 1-µm pitch. Inset: Fourier spectrum of sinusoidal gratings in (b) $y$-direction and (c) $x$-direction. (d) Scanning electron micrograph (SEM; 30° tilt) of the fabricated 40×40 µm$^2$, 150-nm-deep computing layer in silver (Ag). The scalebar is 20 µm. (e,f) Reflected real-space images from flat silver of a projected (e) zero and (f) one. Scale bars are 10 µm. Insets: measured Fourier-space maps showing dominant diffraction into the class-specific predefined output directions. (g) Measured normalized output intensities for all 2,115 test images (980 zeros and 1,135 ones). A correct classification occurs when the designated output channel has the highest intensity. Here, we achieve a classification accuracy of 99.9%. Inset: histogram of classification accuracy per input class.

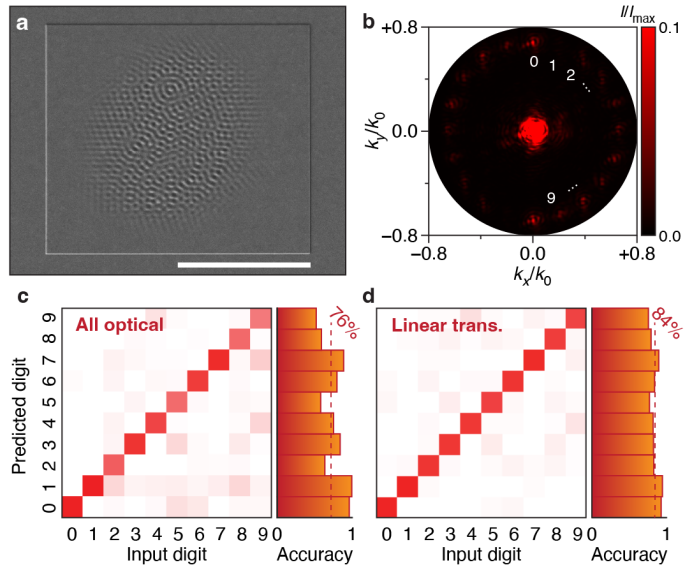

**Figure 2.** Full digit classification. (a) SEM (30° tilt) of the fabricated 40×40 μm², 150-nm-deep diffractive layer, comprising ten class-specific sinusoidal gratings oriented at azimuthal angles incremented by $\pi/10$. The scalebar is 20 μm. (b) Measured Fourier-space map for an input image of a zero, showing dominant diffraction into the two output directions associated with the zero class. Intensity is clipped to $0.1\, I/I_{\max}$ for visibility of the diffraction orders with respect to bright specular reflection. (c) Confusion matrix for all-optical inference across the 10,000 test images. Inset: single-digit accuracies (histogram) and average accuracy $\eta = 76\%$ (dashed line). (d) Confusion matrix and single-digit accuracies after applying an optimized 10×10 linear calibration matrix **M** to the optical outputs. The classification accuracy increases to $\eta = 84\%$.

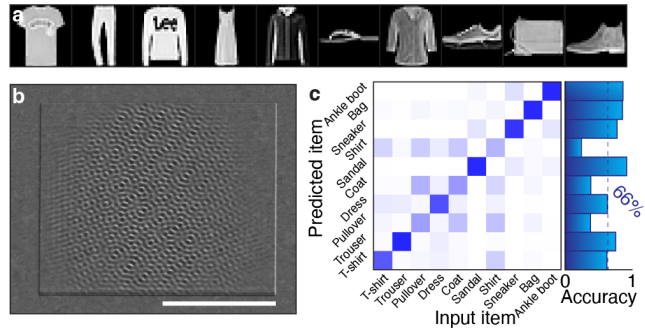

**Figure 3.** More challenging classification tasks. (a) Examples of the 10 classes in the fashion MNIST dataset, comprising visually similar clothing categories that present a more challenging classification task than handwritten digits. (b) SEM (30° tilt) of the diffractive silver layer designed for fashion MNIST classification. The scalebar is 20 μm. (c) Confusion matrix obtained from evaluating all 10,000 fashion MNIST test images using the optically measured outputs combined with the optimized linear calibration matrix **M**. The overall classification accuracy is $\eta = 66\%$. Misclassification primarily occurs among related categories (e.g., shirts, T-shirts, pullovers, coats), reflecting intrinsic dataset ambiguity.

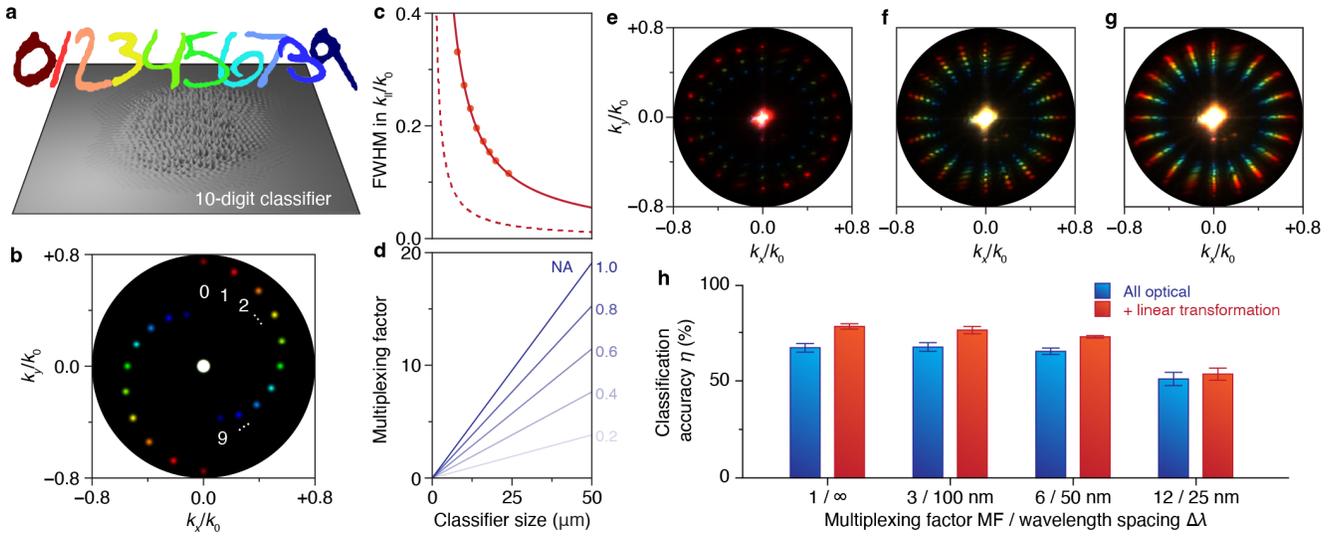

**Figure 4.** Parallel computations using a broadband digit classifier. (a) Concept of wavelength-multiplexed operation: different digit inputs are encoded at different illumination wavelengths, allowing multiple classifications to be processed in parallel by a single diffractive device. (b) Because the diffraction radius in Fourier space scales with wavelength, different colors map to concentric rings, enabling independent readout channels. (c) Calculated full-width-at-half-maximum (FWHM) of the first-order diffraction peaks in Fourier space as a function of classifier size $L$. The dashed line shows the expected $L^{-1}$ scaling for an ideal sinusoidal grating spanning the full device area. Data points indicate the dataset-averaged FWHM obtained by simulating the optical response of 10,000 MNIST training images for different classifier sizes. While the FWHM follows the same $L^{-1}$ scaling (solid line), it exhibits a larger prefactor due to the limited spatial extent of the digit templates within the $L^2$ area. (d) Theoretical multiplexing factor, defined as MF = NA/FWHM, indicating the maximum number of resolvable wavelength channels for a given NA as a function of the classifier size. (e–g) Experimentally measured Fourier-space intensities for multiplexing factors of 3 (450, 550, 650 nm), 6 (425–675 nm in 50 nm steps), and 12 (425–700 nm in 25 nm steps), with all 10,000 test inputs superposed. Diffraction orders remain well separated for 3 and 6 channels, while overlap of diffraction orders for different colors becomes apparent for 12 channels. (h) Average classification accuracy (error bars represent standard deviation) as a function of multiplexing factor MF, shown for all-optical inference (blue) and after application of the linear calibration matrix (red). Robust parallel operation is maintained up to six simultaneous wavelength channels, while higher multiplexing leads to performance degradation because of diffraction-order overlap.

*Supporting Information for*

# Broadband, compact, and training-free optical processors for parallel image classification

*Sander J. W. Vonk, Boris de Jong, Yannik M. Glauser, David B. Seda, Matthieu F. Bidaut, Benjamin Savinson, Hannah Niese, and David J. Norris**

Optical Materials Engineering Laboratory, Department of Mechanical and Process Engineering, ETH Zurich, 8092 Zurich, Switzerland

**Corresponding Author**

*Email: dnorris@ethz.ch

## S1. Theoretical description of optical classification by wavy classifiers

Light diffraction by wavy optical classifiers can be described using scalar diffraction theory, treating the device as a reflective phase plate because of a spatially varying surface height profile. For a more detailed explanation of wavy diffractive surfaces, we refer to Glauser *et al.*[1] Here, we consider a real-valued incoming amplitude field $A(x, y)$ (input image), that is reflected by the optical processor, producing a complex-valued electric field in the device plane

$$E(x, y) = A(x, y) \exp[2i\, k_0\, h(x, y)] \tag{S1}$$

where $h(x, y)$ denotes the surface height profile, $k_0 = 2\pi/\lambda$ is the free-space wave number, and $\lambda$ is the illumination wavelength. The factor of 2 accounts for the reflective geometry, for which the optical path difference is twice the surface height. This electric-field profile propagates away from the device and redistributes the optical energy through diffraction. In the far field (or in Fourier space that we measure in the back focal plane of our microscope objective), we simply find the intensity profile $I_\text{out}$ through the two-dimensional Fourier transform $\mathcal{F}\{\cdot\}$ of the electric field in the device plane, via

$$I_\text{out} \propto |\mathcal{F}\{E\}|^2. \tag{S2}$$

The working principle of optical classification follows from this Fourier-space representation. Each optical classifier consists of a superposition of sinusoidal gratings with a pitch $p = 1$ µm, where each grating is spatially weighted by the average spatial distribution of a specific class. Upon illumination, spatial overlap between the input image $A(x, y)$ and the corresponding class-specific grating template boosts the $\pm 1$ diffraction orders associated with that grating. For a sinusoidal grating of pitch $p$ oriented at an azimuthal angle $\varphi$, the $\pm 1$ diffraction orders appear at well-defined in-plane wavevectors $\mathbf{k}_\parallel = (k_x, k_y)$ given by

$$\frac{\mathbf{k}_\parallel}{k_0} = \pm \frac{\lambda}{p} \begin{pmatrix} \cos \varphi \\ \sin \varphi \end{pmatrix}. \tag{S3}$$

All single-color outputs lie on a circle of radius $k_\parallel/k_0 = \lambda/p$ in Fourier space, with their radial positions determined by the illumination wavelength. These mappings in equation S3 show both the azimuthal angular separation of output channels and the wavelength-multiplexed operation discussed in Figure 4 of the main text.

To simulate the optical response and evaluate classification performance, we numerically compute equation S2 using class-specific illumination patterns $A(x, y)$ derived from the digit MNIST and fashion MNIST datasets. For each input image, the resulting Fourier-space intensity distribution is analyzed to identify the pairs of angularly separated ±1 diffraction orders corresponding to the different output ports. Classification is performed by assigning each input to the class associated with the diffraction channel exhibiting the highest integrated intensity, following the procedure described in equation 2 of the main text. Simulated classification results for digit and fashion item classifiers are summarized as confusion matrices and single-class accuracy histograms in the Supporting Information Figure S3 and Figure S5, respectively.

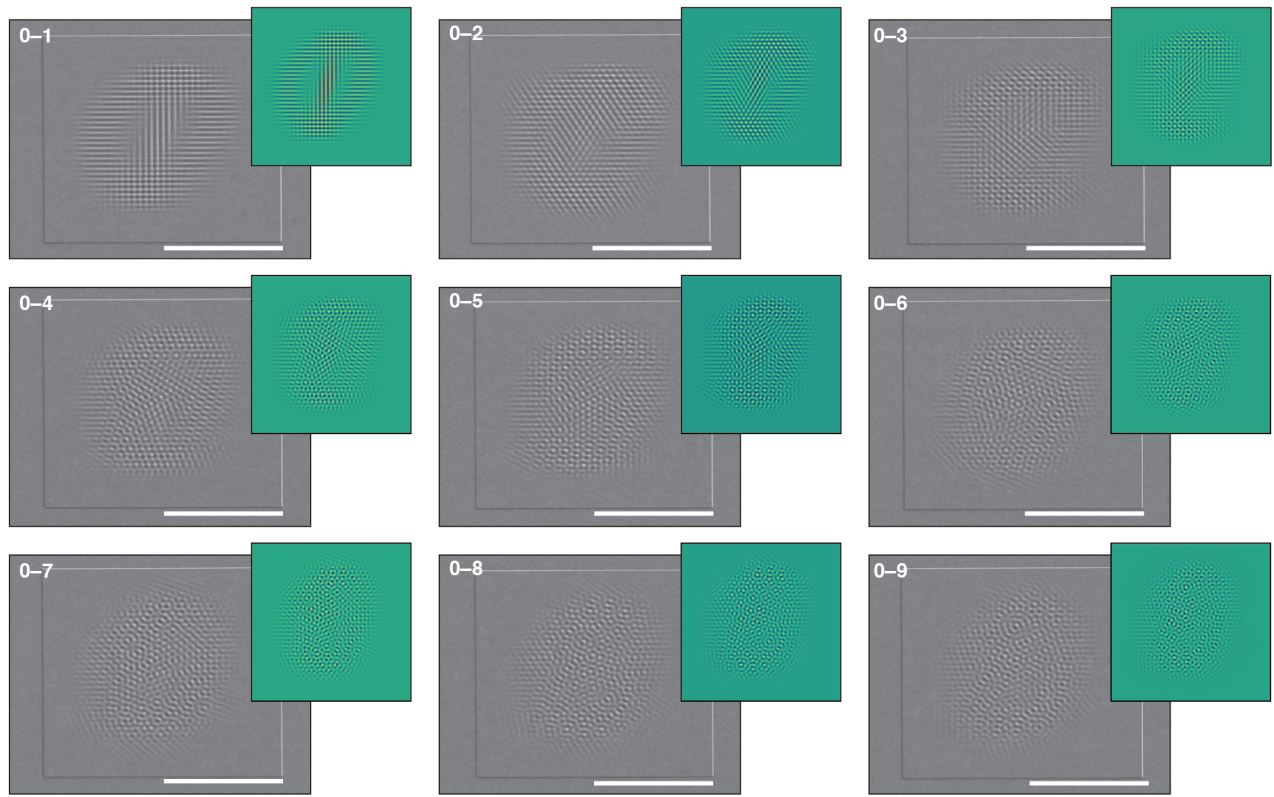

**Figure S1 | Scanning electron micrographs and designs of digit classifiers.** Scanning electron micrographs (SEMs) of all digit classifiers, where the indicated digit range (top left, e.g., 0–9) denotes the set of digit classes contained in each classifier. The number of digit classes was systematically increased, ranging from a 0–1 classifier to a 0–9 classifier. The classifier designs (height profiles) are shown as insets. The scanning electron microscope was operated at 10 kV, and the images were collected at a 30° tilt. All scale bars are 20 μm.

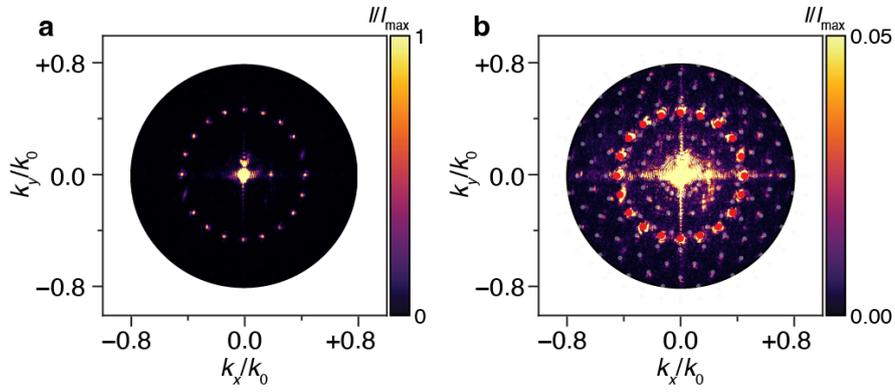

**Figure S2 | Higher-order diffraction. a,** Measured optical response of the full-digit classifier (Figure 2 in the main text), summed over all 10,000 input digits of the MNIST test set. The 20 diffraction spots at $|k_\parallel/k_0| = 0.45$ arise from first-order diffraction of 450-nm illumination. The saturated bright spot at $|k_\parallel/k_0| = 0$ corresponds to specular reflection of the incident light field coming from the surface normal. **b,** Same measurement as in **a**, with the intensity scale clipped to 5% of the maximum intensity. Additional diffraction orders appear next to the first-order diffraction peaks (red dots). These features are attributed to second-order diffraction arising from different combinations of the 10 grating directions, for which the transparent dots indicate the expected diffraction directions. Because the amplitude of each sinusoidal grating is much smaller than the wavelength, second-order diffraction is significantly less efficient than first-order diffraction and therefore has a negligible effect on image classification.

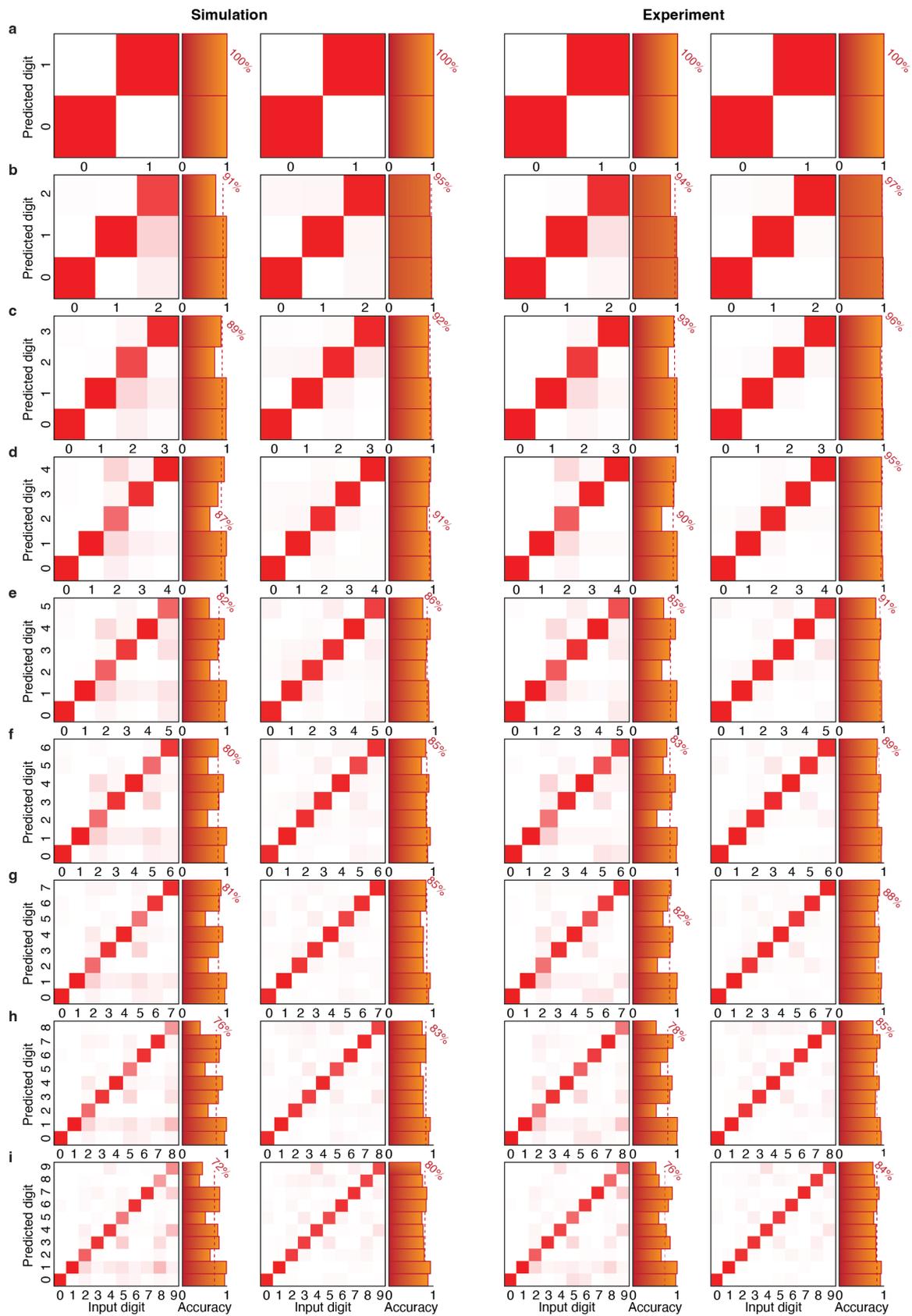

**Figure S3 | Digit classification accuracy. a,** Confusion matrices for binary 0–1 classification: the two left panels are obtained directly from optical outputs (simulated by scalar diffraction theory or measured experimentally), while the two right panels show results after linear transformation of the optical data using an optimized matrix **M**. **b–i,** Same as **a**, but for digit classification from 0–2 up to full 0–9 classification.

S5

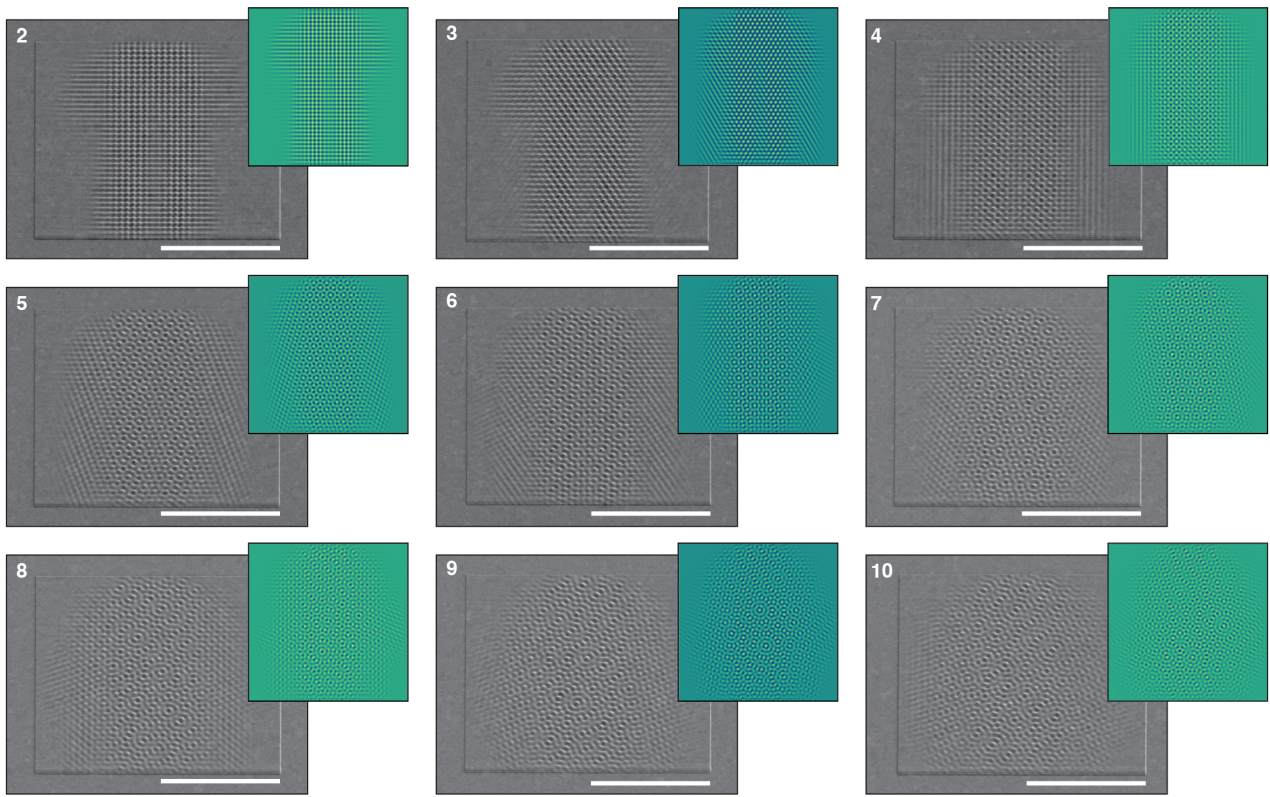

**Figure S4 | Scanning electron micrographs and designs of fashion-item classifiers.** SEMs of all fashion item classifiers, where the indicated digit (top left) denotes the number of fashion classes contained in each classifier. The classifiers always include 'T-shirt' as the first class and progressively incorporate additional classes until all fashion MNIST classes, including 'ankle boot' as the final class, are contained. The classifier designs (height profiles) are shown as insets. The scanning electron microscope was operated at 10 kV, and the images were collected at a 30° tilt. All scale bars are 20 µm.

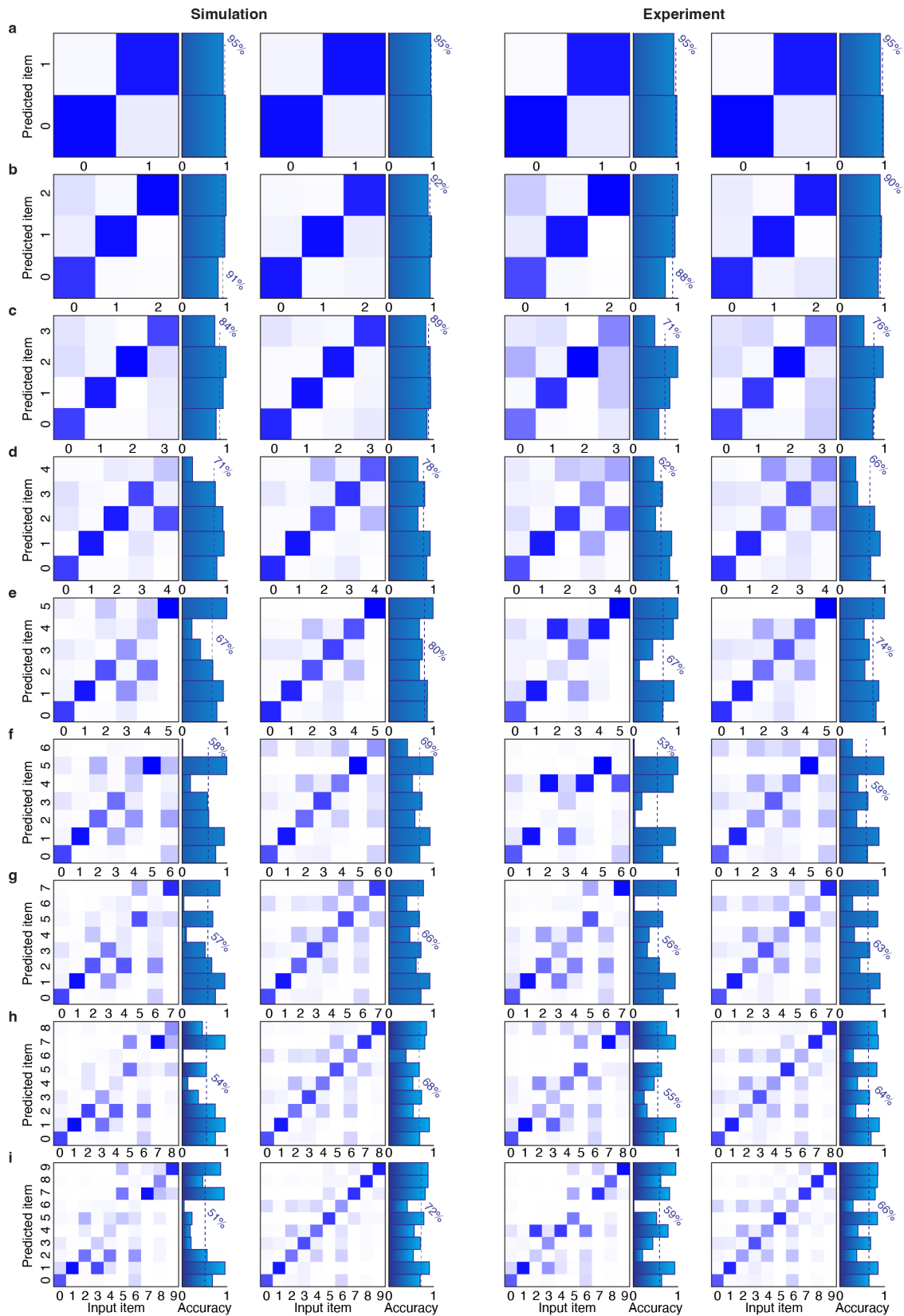

**Figure S5 | Fashion item classification accuracy. a,** Confusion matrices for binary T-shirt–trouser classification: the two left panels are obtained directly from optical outputs (simulated by scalar diffraction theory or measured experimentally), while the two right panels show results after linear transformation of the optical data using an optimized matrix **M**. Fashion-item classes are depicted by a number between 0 (T-shirt) and 9 (ankle boot). **b–i,** Same as **a**, but for fashion-item classification for 3 up to full 10-class classification.

S7

**Supporting references**